\documentclass[12pt,a4paper]{article}
\usepackage[T1]{fontenc}
\usepackage{fancyhdr}
\usepackage{amsmath,amsthm,amsfonts,amssymb}
\allowdisplaybreaks
\usepackage{epic,eepic}
\usepackage{graphicx}
\usepackage{authblk}
\usepackage{fullpage}
\usepackage[active]{srcltx}
\usepackage{enumitem}
\usepackage{booktabs}
\usepackage[left=1.5cm,right=1.5cm]{geometry}
\usepackage{multirow}

\usepackage{framed,color}

\usepackage{float}%
\floatstyle{plaintop}%
\restylefloat{table}

\usepackage[font={small,it}]{caption}

\usepackage[toc,page]{appendix}

\usepackage[style=authoryear,maxcitenames=2,maxbibnames=99]{biblatex}
\bibliography{expSBM}

\usepackage{chngcntr}
\usepackage{dsfont}

\usepackage{algpseudocode}
\usepackage{algorithm}
\usepackage{setspace}

\usepackage{hyperref}

\newcommand{\bmu}{\boldsymbol{\mu}}
\newcommand{\bnu}{\boldsymbol{\nu}}
\newcommand{\blambda}{\boldsymbol{\lambda}}
\newcommand{\bE}{\boldsymbol{\mathcal{E}}}
\newcommand{\btau}{\boldsymbol{\tau}}

\makeatletter
\AtEveryBibitem{%
  \global\undef\bbx@lasthash%
  \clearfield{}}
\makeatother

\theoremstyle{plain}

\newtheorem{proposition}[]{Proposition}

\theoremstyle{definition}

\theoremstyle{remark}

\numberwithin{equation}{section}
\numberwithin{figure}{section}

\date{}
\title{\textbf{A dynamic stochastic blockmodel for interaction lengths}}
\author[1*]{Riccardo Rastelli}
\author[1]{Michael Fop}
\affil[1]{\footnotesize School of Mathematics and Statistics, University College Dublin, Dublin, Ireland}
\affil[*]{\footnotesize riccardo.rastelli@ucd.ie}

\begin{document}
\counterwithout{figure}{section}
\counterwithout{figure}{subsection}
\counterwithout{equation}{section}
\counterwithout{equation}{subsection}

\maketitle
\begin{abstract}
\noindent
We propose a new dynamic stochastic blockmodel that focuses on the analysis of interaction lengths in networks. 
The model does not rely on a discretization of the time dimension and may be used to analyze networks that evolve continuously over time.
The framework relies on a clustering structure on the nodes, whereby two nodes belonging to the same latent group tend to create interactions and non-interactions of similar lengths.
We introduce a fast variational expectation-maximization algorithm to perform inference, and adapt a widely used clustering criterion to perform model choice.
Finally, we test our methodology on artificial data, and propose a demonstration on a dataset concerning face-to-face interactions between students in a high-school.
\\

\end{abstract}

\baselineskip=20pt
\section{Introduction}\label{sec:introduction}
In recent years, a number of network models have been introduced in the literature to study how binary interactions between entities evolve over time. 
One common approach relies on the discretization of the time dimension: once an appropriate time grid is specified, the continuous data are essentially transformed into a collection of static network snapshots.
This approach has facilitated the extension of many static network models to a dynamic framework. 
For example, the Stochastic Block Model (SBM) of \textcite{wang1987stochastic} has been recently adapted to the dynamic case by \textcite{yang2011detecting} and \textcite{matias2017statistical}.
In the same fashion, extensions of the Latent Position Model (LPM) of \textcite{hoff2002latent} have been proposed by \textcite{sarkar2005dynamic} and \textcite{sewell2015latent}, among others.
The model of \textcite{hanneke2010discrete} extends instead the well known Exponential Random Graph Model of \textcite{holland1981exponential}.

However, the approach based on the discretization of the time dimension has been recently criticized, mainly due to the non-negligible effects that the data transformation may have on the results \parencite{corneli2017multiple,matias2018semiparametric}.
In fact, the discretization process always involves a certain level of arbitrariness, either due to the data being collected at specific given times, or because of a post-collection transformation.
In truth, in the vast majority of data analysis applications, the interactions evolve over time in a continuous fashion.

Dynamic binary interactions could either be instantaneous or protracted over a time interval. An example of the first situation is the well-known email network Enron, and this framework has been studied by several recent works, including \textcite{corneli2017multiple} and \textcite{matias2018semiparametric}. However, in many situations the interactions among a collection of entities may be protracted over time, and the object of analysis may be to model for how long these entities interact (and conversely do not interact) within an observed time period: in this paper we focus on such case.
This context generates data which allow a representation of the generic interaction using the format $(i,j,t,\ell)$, where $i$ denotes the \textit{sender} node, $j$ the \textit{receiver}, $t$ is the instant in which the interaction begins, and $\ell$ the interaction length. 
This framework is apt to describe a variety of networks, including phone call networks, visual contact networks, speech networks, or proximity networks.

The goal of this paper is to introduce a continuous network framework to directly model the lengths of the observed binary interactions.
Our approach relies on a SBM structure, whereby the nodes are characterized by a cluster membership variable (\textit{allocation}), which determines the distributions over the edge values.
In our context, the allocations determine both the lengths of the interactions as well as the lengths of the non-interactions.

The literature on the modeling of interaction lengths is rather limited.
One framework that is similar to ours is the Stochastic Block Transition Model of \textcite{xu2015stochastic} and \textcite{rastelli2018exact}.
In these papers, the block structure is used to determine the propensity to create and destroy edges across contiguous time frames.
This can naturally give a model-based quantification of the persistence of edges and non-edges.
Our work also shares similarities with the Stochastic Actor-Oriented Models, discussed by \textcite{snijders2005models}, and extended towards a number of different directions in more recent contributions.
We should point out that, while these works have motives and methods similar to ours, both of these approaches rely on the discretization of the time dimension, whereas our proposed model is based on a fully continuous framework.

As regards model inference, we propose a variational expectation-maximization algorithm to estimate model parameters and cluster allocation. Additionally, a model-based clustering criterion is introduced to select the optimal number of latent groups.
In recent years, variational methods have been successfully applied in a variety of mixture models for networks.
For example, they have been employed for the static and dynamic SBM by \textcite{daudin2008mixture} and \textcite{matias2017statistical}, respectively.
They have also been used for mixed membership models \parencite{airoldi2008mixed}, networks for instantaneous interactions \parencite{corneli2017multiple}, and textual networks \parencite{bouveyron2018stochastic}.
A recent review on variational inference can be found in \textcite{blei2017variational}.

The paper is organized as follows: Section \ref{sec:model} describes in detail the type of data analyzed, introducing a homogeneous model and its new stochastic block model extension.
Section \ref{sec:inference} presents a variational expectation-maximization algorithm to estimate the model parameters, and a criterion to select the number of clusters.
In Section \ref{sec:sim} the proposed method is tested on simulated data experiments, whereas in Section \ref{sec:data} it is demonstrated in application to the analysis of face-to-face interactions between high-school students in France. The paper ends with a discussion in \ref{sec:conclusions}.

\section{Statistical model}\label{sec:model}
\subsection{Interaction length data}\label{sec:interaction_length_data}
The observed data describe the binary interactions between units during a certain time interval $[0,T]$.
These units, or nodes, are labeled with $\mathcal{N} = \left\{1,\dots,N\right\}$.
At every point in time $t\in[0,T]$, node $i\in\mathcal{N}$ may or may not be interacting with node $j\in\mathcal{N}$.
We illustrate our model focusing on directed interactions, noting that extensions to the undirected case are straightforward to implement.

We represent the data observations with a continuous collection of adjacency matrices 
\begin{equation}
 \boldsymbol{\mathcal{E}} = \left\{ \mathcal{E}_{ij}(t) \in \{0,1\},\ \forall t \in [0,T],\ \forall i\in\mathcal{N},\ \forall j\in\mathcal{N},\ i\neq j\right\},
\end{equation}
where $\mathcal{E}_{ij}(t)$ is equal to $1$ whenever $i$ is interacting with $j$, and to $0$ otherwise.

Since $\mathcal{E}_{ij}(\cdot)$ is a step function, we can avoid the continuous notation.
We note that, for each pair of nodes, $\mathcal{E}_{ij}(\cdot)$ naturally partitions the segment $[0,T]$ separating the subsets where the function takes the same value.
We denote these subsets with $\mathcal{E}_{ij}^{(1)}, \mathcal{E}_{ij}^{(2)},\dots$, and impose two conditions: 
\begin{equation}
    \begin{split}
        &\forall w=1,\ 2,\ \dots \colon\\
        &\hspace{1cm}\mathcal{E}_{ij}(t_1) = \mathcal{E}_{ij}(t_2) \hspace{1cm} \forall t_1, t_2 \in \mathcal{E}_{ij}^{(w)} \\
        &\hspace{1cm}\mathcal{E}_{ij}(t_1) \neq \mathcal{E}_{ij}(t_2) \hspace{1cm} \forall t_1 \in \mathcal{E}_{ij}^{(w)},\ \forall t_2 \in \mathcal{E}_{ij}^{(w+1)}
    \end{split}
\end{equation}
In other words, $\mathcal{E}_{ij}(\cdot)$ must be constant on each set, and any two consecutive sets cannot correspond to the same value of $\mathcal{E}_{ij}(\cdot)$. In practice, the data must be represented as an alternating sequence of interactions and non-interactions.

In particular, we are interested in studying the length of these interactions, which we denote $X_{ij}^{(w)}$, for $w=1,2,\dots$.
Also, we denote with $A_{ij}^{(w)}$ the value of $\mathcal{E}_{ij}(\cdot)$ on the corresponding set.
Exploiting the fact that $\mathcal{E}_{ij}(\cdot)$ can only take values $0$ and $1$, we can actually reconstruct all of the observed data just by using the collection of $\left\{X_{ij}^{(w)}\right\}_{i,j,w}$ and the initial values $\left\{A_{ij}^{(1)}\right\}_{i,j}$.

\subsection{The homogeneous model}\label{sec:homogeneous_model}
From now on, all the probabilities considered are conditional on $T$ and the initial states $\left\{A_{ij}^{(1)}\right\}_{i,j}$. 

We assume that each interaction length (resp. non-interaction length) is an independent exponential random variable with rate $\mu$ (resp. $\nu$), hence the value $1/\mu$ denotes the average time of interaction (resp. $1/\nu$ the average time of non-interaction).
Since these rates do not depend on the nodes at the extremities of each edge, the model is homogeneous.

\begin{figure}[htbp]
    \centering
    \includegraphics{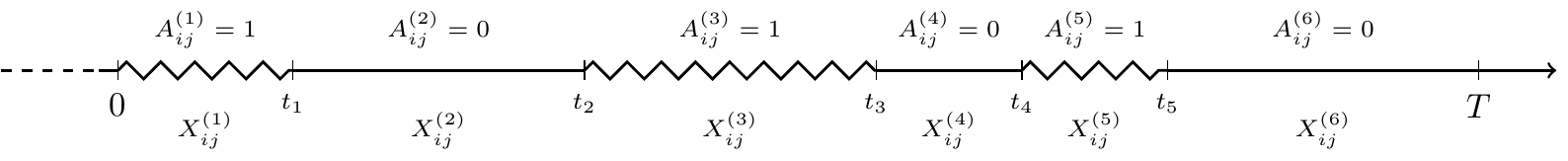}
    \caption{A graphical representation of the data provided by a generic edge $(i,j)$. In this case, since $W_{ij} = 6$, there are $4$ embedded sub-segments which yield the exponential interaction lengths $X_{ij}^{(3)}$ and $X_{ij}^{(5)}$, and the non-interaction lengths $X_{ij}^{(2)}$ and $X_{ij}^{(4)}$. The interaction length $X_{ij}^{(1)}$ and non-interaction length $X_{ij}^{(6)}$ are truncated from the left and from the right, respectively.}
    \label{fig:edge_graphical_representation}
\end{figure}

Consider the edge between two arbitrary nodes $i\in\mathcal{N}$ and $j\in\mathcal{N}$.
The interval $[0,T]$ is partitioned into the sets $\mathcal{E}_{ij}^{(1)}, \dots, \mathcal{E}_{ij}^{(W_{ij})}$, where $W_{ij}$ denotes the total number of interactions and non-interactions between $i$ and $j$.
One important aspect of these data is that the interaction (or non-interaction) lengths $X_{ij}^{(1)}$ and $X_{ij}^{(W_{ij})}$ only provide a lower bound for the true non-observed lengths.
In other words, these observed values are truncated.
A representation of the data structure under consideration is presented in Figure \ref{fig:edge_graphical_representation}.

It follows that the likelihood contribution provided by an edge with more than $2$ sub-segments is given by a product of $W_{ij}-2$ exponential densities, corresponding to the observed embedded intervals, and two cumulated densities, corresponding to the truncated observations at the extremities. 
If the number of segments is equal to $2$, only the cumulated densities must remain.
If the number of segments is equal to $1$, only one cumulated density must remain, which, using the exponential assumption, can be imposed with $X_{ij}^{w} = 0$ for all $w>1$.

%
%
These properties translate into the following probability for the pair of nodes $(i,j)$:
\begin{equation}
\begin{split}
p\left( \boldsymbol{\mathcal{E}}_{ij} \middle\vert \mu,\nu \right) 
&= \left[ 1-F_\mu\left( X_{ij}^{(1)} \right)\right]^{A_{ij}^{(1)}}\left[ 1-F_\nu\left( X_{ij}^{(1)} \right)\right]^{1-A_{ij}^{(1)}} \\
&\hspace{1cm}\times\prod_{w=2}^{W_{ij}-1} f_\mu\left( X_{ij}^{(w)} \right)^{A_{ij}^{(w)}}f_\nu\left( X_{ij}^{(w)} \right)^{1-A_{ij}^{(w)}} \\
&\hspace{1cm}\times\left[ 1-F_\mu\left( X_{ij}^{(W_{ij})} \right)\right]^{A_{ij}^{(W_{ij})}}\left[ 1-F_\nu\left( X_{ij}^{(W_{ij})} \right)\right]^{1-A_{ij}^{(W_{ij})}}
\end{split}
\end{equation}
where $f_\lambda(\cdot)$ and $F_\lambda(\cdot)$ are the pdf and cdf of an exponential variable with rate $\lambda$, respectively.

To summarize the observed data, we can introduce the following statistics:
\begin{equation}
    \begin{split}
        \mathcal{A}_{ij}^{(+)} &= \sum_{w=2}^{W_{ij}-1} A_{ij}^{(w)} \hspace{2cm} \mathcal{A}_{ij}^{(-)} =  \sum_{w=2}^{W_{ij}-1} \left(1-A_{ij}^{(w)}\right) \\
        \mathcal{X}_{ij}^{(+)} &= \sum_{w=1}^{W_{ij}} X_{ij}^{(w)}A_{ij}^{(w)} \hspace{2cm} \mathcal{X}_{ij}^{(-)} = \sum_{w=1}^{W_{ij}} X_{ij}^{(w)}\left(1-A_{ij}^{(w)}\right)
    \end{split}
\end{equation}

Now, we replace the exponential distributions with their actual expressions, and, using the independence assumption on the edges, we obtain the following log-likelihood:
\begin{equation}
 \begin{split}
\ell_{\boldsymbol{\mathcal{E}}}\left( \mu, \nu \right) 
= \sum_{i \neq j} \log p\left( \boldsymbol{\mathcal{E}}_{ij} \middle\vert \mu,\nu \right) = L_{\mu} \log \left( \mu \right) + L_{\nu} \log \left( \nu \right) - \mu\eta - \nu\zeta
 \end{split}
\end{equation}
where:
\begin{equation}
\begin{split}
 L_{\mu} &=  \sum_{i \neq j} \mathcal{A}_{ij}^{(+)} \hspace{2cm}
 L_{\nu} = \sum_{i \neq j} \mathcal{A}_{ij}^{(-)} \\
 \eta &= \sum_{i \neq j} \mathcal{X}_{ij}^{(+)} \hspace{2cm}
 \zeta = \sum_{i \neq j} \mathcal{X}_{ij}^{(-)} \\
\end{split}
\end{equation}

Maximum likelihood estimators are available in closed form for the two model parameters:
\begin{equation}
 \hat{\mu} = \frac{L_{\mu}}{\eta} \hspace{2cm}
 \hat{\nu} = \frac{L_{\nu}}{\zeta}
\end{equation}

\subsection{The Stochastic Blockmodel for interaction lengths}\label{sec:blockmodel}
A natural extension of the homogeneous model is a latent block structure model, where nodes are allowed to belong to different sub-populations.
An allocation variable $Z_i$ is thus assigned to each of the nodes, to indicate their cluster membership. Such categorical variable $Z_i$ takes values in $\mathcal{K} = \left\{ 1,\dots,K\right\}$ and indicates to which of $K$ groups node $i\in\mathcal{N}$ is allocated to. 
As in any finite mixture model, these allocation variables are assumed to arise from a Multinomial distribution with probabilities $\blambda = \left\{\lambda_1,\dots,\lambda_K\right\}$, where the generic $\lambda_k$ corresponds to probability of observing a priori group $k$.
Since the number of groups must be chosen and fixed, a convenient alternative representation is given by $Z_{ik} = \mathds{1}_{\left\{ Z_i=k\right\}}$, where $\mathds{1}_{\mathcal{S}}$ denotes the indicator function for the set $\mathcal{S}$.
Finally, the blockmodel assumption postulates that the lengths of the interactions and the lengths of non-interactions between a node in group $g$ and a node in group $h$ are determined by the parameters $\mu_{gh}$ and $\nu_{gh}$, respectively.

The conditional log-likelihood of the SBM reads as follows:
\begin{equation}\label{eq:conditional_likelihood_1}
 \begin{split}
  \ell_{\bE}\left( \bmu, \bnu, \textbf{Z} \right) 
 &= \log p\left( \bE \middle\vert \bmu, \bnu, \textbf{Z} \right)
 = \sum_{i\neq j} \sum_{g=1}^K \sum_{h=1}^K Z_{ig} Z_{jh} \log p\left( \bE_{ij} \middle\vert \mu_{gh},\nu_{gh} \right)
 \end{split}
\end{equation}
This formulation mimics the finite mixture framework in the network context, as in \textcite{daudin2008mixture}.
We can follow the same procedure used for the homogeneous model to obtain the following result:
\begin{equation}\label{eq:conditional_likelihood_2}
 \begin{split}
  \ell_{\bE}\left( \bmu, \bnu, \textbf{Z} \right) 
&= \sum_{g=1}^K \sum_{h=1}^K \left\{L_{\mu_{gh}} \log \left( \mu_{gh} \right) + L_{\nu_{gh}} \log \left( \nu_{gh} \right) - \mu_{gh}\eta_{gh} - \nu_{gh}\zeta_{gh}\right\}
 \end{split}
\end{equation}
where the new quantities are defined as:
\begin{equation}
\begin{split}
 L_{\mu_{gh}} &=  \sum_{i \neq j} Z_{ig} Z_{jh} \mathcal{A}_{ij}^{(+)} \hspace{1.5cm}
 L_{\nu_{gh}} = \sum_{i \neq j} Z_{ig} Z_{jh} \mathcal{A}_{ij}^{(-)} \\
 \eta_{gh} &= \sum_{i \neq j} Z_{ig} Z_{jh} \mathcal{X}_{ij}^{(+)} \hspace{1.5cm}
 \zeta_{gh} = \sum_{i \neq j} Z_{ig} Z_{jh} \mathcal{X}_{ij}^{(-)} \\
\end{split}
\end{equation}

\section{Inference}\label{sec:inference}
\subsection{Variational Expectation-Maximization algorithm}\label{sec:VEM}
As is usual in model-based clustering, we are interested in performing inference for this model by maximizing the marginal likelihood (or evidence) $p\left( \bE \middle\vert \blambda, \bmu, \bnu \right)$ with respect to the model parameters. 
However, integrating out the allocations $\textbf{Z}$, is not computationally feasible, even for very small datasets.
An Expectation-Maximization (EM) algorithm \parencite{dempster1977maximum} can be employed to overcome this issue.

The EM alternates two steps: the E-step, where we calculate the expectation of the conditional likelihood \eqref{eq:conditional_likelihood_1} with respect to the posterior $\pi\left( \textbf{Z} \middle\vert \bE \right)$, and the M-step, where we maximize such expectation with respect to the likelihood parameters.
The combination of the two steps is guaranteed to not decrease the value of the objective function (in this case, the marginal likelihood) and to converge to a local optimum \parencite{wu:1983}. However, differently from other more common finite mixture models, the posterior distribution $\pi\left( \textbf{Z} \middle\vert \bE \right)$ does not factorize into a simple form for stochastic blockmodels such as ours.
As a consequence, the E-step cannot be performed exactly, due to the higher computational costs.
This makes the standard EM algorithm not applicable.

A variational approximation can be used to overcome this limitation and perform inference on SBM, as previously proposed by \textcite{daudin2008mixture} and a number of subsequent works.
The goal here is to replace the posterior distribution on the allocations by a more tractable one, which would allow an efficient use of the EM algorithm.
In practice, we introduce variational parameters $\btau$ and consider the family of all the distributions $q$ that factorize into the product of independent multinomial variables:
\begin{equation}
 q\left( \textbf{Z} \middle \vert \btau \right)
 = \prod_{i=1}^N q\left( \textbf{Z}_{i} \middle \vert \tau_{i1},\dots,\tau_{iK}\right) 
 = \prod_{i=1}^N \prod_{k=1}^K \tau_{ik}^{Z_{ik}} 
\end{equation}
then, the distribution that is most similar to the true posterior $\pi$ is selected.

This can be formalized as follows.
First, we note that the marginal log-likelihood satisfies:
\begin{equation}
 \log p\left( \bE \middle \vert \bmu, \bnu,  \blambda\right)
 = \log p\left( \bE, \textbf{Z} \middle \vert \bmu, \bnu,  \blambda \right) - \log \pi\left( \textbf{Z} \middle \vert \bE, \bmu, \bnu,  \blambda \right)
\end{equation}
Now, we add and subtract $\log q\left( \textbf{Z} \middle \vert \btau \right)$ on the right hand side, and take the expectations of both sides, obtaining:
\begin{equation}\label{eq:elbo_1}
 \log p\left( \bE \middle \vert \bmu, \bnu, \blambda \right)
 = \mathcal{KL}\left( q || \pi  \right) + \mathbb{E}_q\left[ \log p\left( \bE, \textbf{Z} \middle \vert \bmu, \bnu,  \blambda \right) \right] + Ent(q)
\end{equation}
Here, $\mathcal{KL}$ refers to the Kullback-Leibler divergence and is defined as:
\begin{equation}
 \mathcal{KL}\left( q || \pi \right) = \mathbb{E}_q\left[ \log q\left( \textbf{Z} \middle \vert \btau \right) - \log \pi\left( \textbf{Z} \middle \vert \bE, \bmu, \bnu, \blambda \right) \right]
\end{equation}
whereas the entropy of the variational distribution is defined as:
\begin{equation}
 Ent(q) = - \mathbb{E}_q\left[ \log q\left( \textbf{Z} \middle \vert \btau \right) \right]
\end{equation}

The decomposition of the marginal log-likelihood in \eqref{eq:elbo_1} can be exploited to define an optimization procedure.
We alternate two steps: in the first step, we minimize $\mathcal{KL}\left( q || \pi \right)$ with respect to the variational parameters $\btau$; 
in the second step, we maximize the right hand side of \eqref{eq:elbo_1} with respect to the likelihood parameters.

In practice, the two steps correspond to the maximization of the lower bound to the marginal likelihood given by 
$\mathcal{F} = \mathbb{E}_q\left[ \log p\left( \bE, \textbf{Z} \middle \vert \bmu, \bnu,  \blambda \right) \right] + Ent(q)$ with respect to $\left\{ \btau \right\}$ and $\left\{ \blambda, \bmu, \bnu \right\}$, respectively.
In our SBM for interaction lengths, the quantities involved in \eqref{eq:elbo_1} can be written down explicitly, and the update rules are determined by closed form equations, as shown in the next section.

Once the algorithm stops, it returns the optimal parameters $\hat{\bmu}$, $\hat{\bnu}$, $\hat{\blambda}$ and $\hat{\btau}$.
Regarding the clustering task, the parameters $\hat{\btau}$ denote a soft partition for the nodes and represents an approximation to the allocation variable $\mathbf{Z}$.
These values may be interpreted as the posterior probabilities for the nodes to belong to each of the $K$ groups. Hence, a straightforward estimated hard partition $\hat{\textbf{Z}}$ can simply be obtained by considering the maximum a posteriori derived from $\hat{\btau}$.

\subsection{Update rules}\label{sec:formulas}
First, we characterize our objective function with the following proposition.
\begin{proposition}\label{prop:elbo_1}
The evidence lower bound for the model proposed is given by:
\begin{equation}\label{prop:elbo_1_eq_1}
\begin{split}
 \mathcal{F} &= \sum_{g=1}^K \sum_{h=1}^K \left\{\bar{L}_{\mu_{gh}} \log \left( \mu_{gh} \right) + \bar{L}_{\nu_{gh}} \log \left( \nu_{gh} \right) - \mu_{gh}\bar{\eta}_{gh} - \nu_{gh}\bar{\zeta}_{gh}\right\} \\
 &\hspace{1cm}+ \sum_{i=1}^N \sum_{k=1}^K \tau_{ik} \log \lambda_k - \sum_{i=1}^N \sum_{k=1}^K \tau_{ik} \log \tau_{ik} 
\end{split}
\end{equation}
where the tildes denote the expected values of the corresponding quantities with respect to $q$:
\begin{equation}
\begin{split}
 \bar{L}_{\mu_{gh}} &=  \sum_{i \neq j} \tau_{ig} \tau_{jh} \mathcal{A}_{ij}^{(+)} \hspace{1.5cm}
 \bar{L}_{\nu_{gh}} = \sum_{i \neq j} \tau_{ig} \tau_{jh} \mathcal{A}_{ij}^{(-)} \\
 \bar{\eta}_{gh} &= \sum_{i \neq j} \tau_{ig} \tau_{jh} \mathcal{X}_{ij}^{(+)} \hspace{1.5cm}
 \bar{\zeta}_{gh} = \sum_{i \neq j} \tau_{ig} \tau_{jh} \mathcal{X}_{ij}^{(-)}
\end{split}
\end{equation}
\end{proposition}
The proof is given in Appendix \ref{app:elbo_1}.

The following propositions focus instead on the optimization of the evidence lower bound with respect to the variational parameters and the model parameters.
\begin{proposition}\label{prop:elbo_tau}
The variational parameters that minimize $\mathcal{KL}\left(q||\pi\right)$ are given by:
\begin{equation}\label{prop:elbo_tau_eq_1}
\begin{split}
 \hat{\tau}_{\ell k} = \frac{ \sum_{j=1}^N \sum_{h=1}^K \tau_{jh}\omega_{\ell jkh} + \sum_{i=1}^N \sum_{g=1}^K \tau_{ig}\omega_{i\ell gk} + \log \lambda_k}
 {\sum_{k=1}^K \left\{\sum_{j=1}^N \sum_{h=1}^K \tau_{jh}\omega_{\ell jkh} + \sum_{i=1}^N \sum_{g=1}^K \tau_{ig}\omega_{i\ell gk} + \log \lambda_k \right\}} 
\end{split}
\end{equation}
where:
\begin{equation}\label{prop:elbo_tau_eq_2}
\begin{split}
\omega_{ijgh} 
= \mathcal{A}_{ij}^{(+)} \log\left(\mu_{gh}\right) + \mathcal{A}_{ij}^{(-)} \log\left(\nu_{gh}\right) - \mathcal{X}_{ij}^{(+)} \mu_{gh} - \mathcal{X}_{ij}^{(-)} \nu_{gh}\end{split}
\end{equation}
\end{proposition}
The proof is given in Appendix \ref{app:elbo_tau}.

\begin{proposition}\label{prop:elbo_lambda}
The optimal mixing proportions are given by:
\begin{equation}
  \hat{\lambda}_k = \frac{ \sum_{i=1}^N \tau_{ik} }{N}
\end{equation}
\end{proposition}
The proof is given in Appendix \ref{app:elbo_lambda}.

\begin{proposition}\label{prop:elbo_munu}
The model parameters maximizing the lower bound for the evidence are:
\begin{equation}
 \begin{split}
 \hat{\mu}_{gh} = \frac{\bar{L}_{\mu_{gh}}}{\bar{\eta}_{gh}} \hspace{2cm}
 \hat{\nu}_{gh} = \frac{\bar{L}_{\nu_{gh}}}{\bar{\zeta}_{gh}}
 \end{split}
\end{equation}
\end{proposition}
The proof is given in Appendix \ref{app:elbo_munu}.

\subsection{Algorithm initialization}\label{sec:initialization}
For a fixed value of $K$, the variational EM algorithm allows one to perform inference on the model parameters $\bmu, \bnu,  \blambda$, and cluster allocations $\textbf{Z}$. The procedure needs to be initialized from some starting values as the algorithm is only guaranteed to converge to a local optimum, and often, in the context of mixture models, the final estimates are dependent on such initial values \parencite{baudry:2015,ohagan:2012,scrucca:2015,biernacki:2003}. Several strategies have been proposed in the literature in the context of SBM for networks, see \textcite{come2015model,matias2017statistical,bouveyron2018stochastic}, for example. 
A simple strategy is to consider multiple random starting allocations and then retain the model with the highest value of the maximized lower bound of the marginal likelihood as described in Section~\ref{sec:VEM}. However, a random initialization does not avoid that the algorithm could reach a sub-optimum solution and is often computationally intensive \parencite{scrucca:2015,come2015model}. Hence, we adopt the following initialization procedure based on spectral clustering in order to provide an initial estimate of the allocations $\mathbf{Z}$:
\begin{enumerate}[noitemsep]
    \item Compute the total interaction duration time between any pair of nodes $(i,j)$ and construct the $N\times N$ matrix $\mathbf{M}$.  
    \item Perform spectral $K$-means clustering \parencite{von:2007} using the affinity matrix $\log(\mathbf{M} + \mathbf{M}^{\top})/2$, where the logarithm is taken element-wise and only for non-zero entries.
    \item Initialize the allocations $\mathbf{Z}$ using the classification obtained in the spectral clustering step.
\end{enumerate}
The rationale of this procedure is that nodes interacting more often and for longer time are reasonably expected to belong to the same cluster, and that the affinity matrix constructed using the total interaction time over the observed period would naturally include this information.

\subsection{Model selection}\label{sec:icl}
The optimal number of latent groups $K$ is often not known and needs be estimated from the data: we propose to choose the value of $K$ that maximizes the Integrated Completed Likelihood (ICL) criterion.
The ICL criterion, first introduced by \textcite{biernacki2000assessing}, aims at maximizing the integrated completed likelihood $p\left( \bE, \textbf{Z} \middle\vert \blambda, \bmu, \bnu \right)$.
This criterion has been widely used to perform model choice for mixture models, especially within the literature on networks \parencite{daudin2008mixture,come2015model,rastelli2018choosing}.

We propose to evaluate this criterion using a BIC-type approximation, as previously proposed by several other works, including \textcite{biernacki2000assessing}, \textcite{daudin2008mixture} and \textcite{matias2017statistical}.
\begin{proposition}\label{prop:icl_1}
In our proposed model, the ICL value is equal to:
\begin{equation}
    ICL_{ex}(K) = \max_{\blambda,\bmu,\bnu}\log p\left( \bE, \hat{\textbf{Z}} \middle\vert \blambda,\bmu,\bnu \right) - K^2 \log \left( \sum_{i\neq j} W_{ij} \right) - \frac{K-1}{2} \log N
\end{equation}
where $\hat{\textbf{Z}}$ denotes the Maximum-A-Posteriori estimates of the allocations, as obtained by the variational EM algorithm.
\end{proposition}
The proof of this proposition is straightforward, since the formula is simply an adaptation of a similar result from \textcite{daudin2008mixture}.

We note that, in practice, the ICL criterion requires fitting the model for every plausible value of $K$.
As the computational cost of the variational EM grows with $K$, this grid search may slow down the procedure unnecessarily.
For this reason, we generally consider values of $K$ that are much smaller than $N$, run the variational EM for those, and select the best result using the ICL values.

\section{Simulated data experiments}\label{sec:sim}
We propose three simulation studies to assess the performance of our method with respect to clustering and model selection.

\subsection{Simulation study 1}
In the first study, our goal is to assess the performance of our method in clustering the nodes.
We generate $100$ random networks of $100$ nodes from our likelihood model. 
We consider $K=3$ latent groups.
The true mixing proportions $\blambda$ are generated independently for each network using a symmetric Dirichlet distribution with parameter $0.5$.
For each of the latent groups, the rates $\bmu$ and $\bnu$ are generated independently from a gamma distribution with shape and rate both equal to $\xi$.
This distribution has mean $1$ and variance equal to $1/\xi$.
In other words, $\xi$ is the precision associated to this gamma distribution and determines whether the groups are well separated or not.
A small $\xi$ value will imply an easier inferential task, where clusters are well separated; vice-versa, as $\xi$ increases, separating the cluster will become more challenging.
We repeat the experiment for every value of $\xi$ in the set $\left\{0.5, 1, 5, 25, 50\right\}$.

Our model assumes that, for each pair of nodes, their first and last interactions are truncated. 
In order to mimic this behavior, we propose the following mechanism.
First, for every pair of nodes, we sample the values $\left\{A_{ij}^{(1)}\right\}_{i,j}$ uniformly at random from the set $\left\{0,1\right\}$.
Then, we consider a time horizon $T=10$, and, for all pairs of nodes, we generate a sequence $x^{(1)},x^{(2)},\dots$ of i.i.d. exponential random variables, using the appropriate exponential rates, as indicated by the derived values of $A_{ij}^{(w)}$ and by cluster memberships.
Finally, we truncate the sequence at the $W$-th value such that $\sum_{w=1}^{W-1} x^{(w)} < T$ and $\sum_{w=1}^W x^{(w)} \geq T$. 
Crucially, we truncate the last generated value to make sure that the overall length is exactly equal to $T$.
In this way, every pair of nodes have cumulated interaction and non-interaction lengths equal to $T$.

\begin{figure}[ht]
\centering
\includegraphics[width=0.49\textwidth]{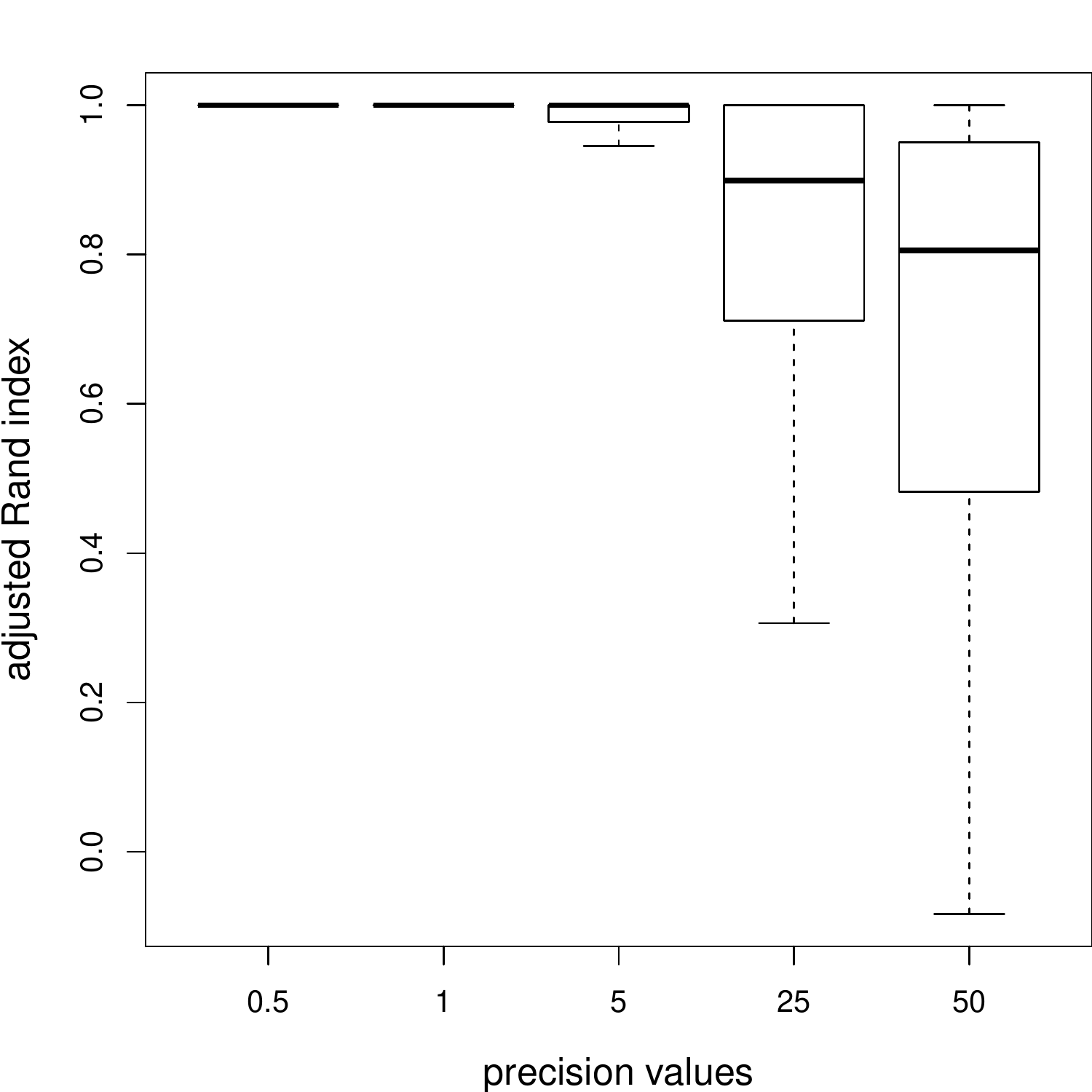}
\includegraphics[width=0.49\textwidth]{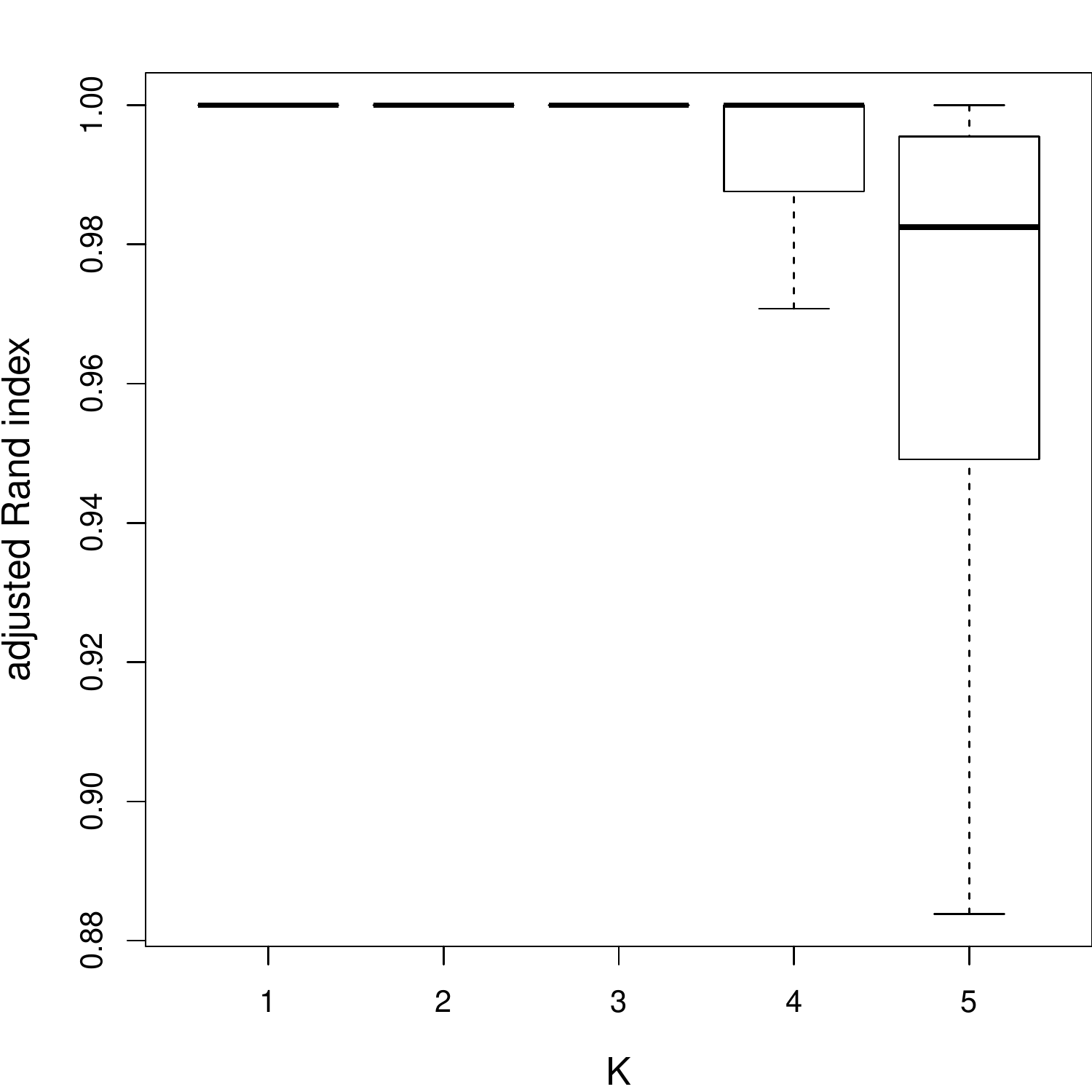}
\caption{\textbf{Simulation studies 1 and 2}. Adjusted Rand index between the true partition and the estimated maximum a posteriori partition, for the first and second simulation studies on the left and right panel, respectively. Note the different scales on the vertical axis.}
 \label{fig:sim_1a_2a}
\end{figure}

We use our inferential procedure once on each dataset, providing in input the correct number of groups to the algorithm.
The computing times are reported in Table \ref{tab:sim_1a}.
\begin{table}[bt]
\centering
\begin{tabular}{cccccc}  
\toprule
&\multicolumn{5}{c}{$\xi$} \\
\cmidrule(r){2-6}
& 0.5 & 1 & 5 & 25 & 50 \\
\midrule
Seconds & 0.021 & 0.019 & 0.014 & 0.013 & 0.013 \\
\bottomrule
\end{tabular}
\caption{\textbf{Simulation study 1}. Average number of seconds for each of the values of $\xi$ considered.}
\label{tab:sim_1a}
\end{table}
We assess the clustering performance by comparing the true and estimated partitions using the adjusted Rand index (ARI) introduced by \textcite{hubert1985comparing}. 
In particular, we use the estimated maximum a posteriori partition $\hat{\textbf{Z}}$.
The left panel of Figure \ref{fig:sim_1a_2a} shows summaries for the obtained ARI values for each of the $\xi$ values considered.
The smallest two values of $\xi$ lead to well separated clusters, which are always successfully identified by our method.
In these cases there are no differences between the true and estimated partitions.
As $\xi$ increases, the algorithm maintains a good performance in most of the generated datasets, however it fails to recognize the groups in some of the networks.

\subsection{Simulation study 2}
In the second simulation study, our goal is to assess the performance of our method in both clustering and model choice.
We consider again $100$ artificial networks of $100$ nodes.
We repeat the experiment five times, i.e. for the true $K$ varying in the set $\left\{1,2,3,4,5\right\}$.
In this study, we do not use the parameter $\xi$: we consider instead a community structure, where nodes belonging to the same group tend to interact more frequently and for a longer time.
For each value of $K>1$, the matrix $\bmu$ has values $\varepsilon = 0.5$ on the diagonal and $\theta = 5$ on the off-diagonal elements. 
Viceversa, the matrix $\bnu$ has values $\theta$ on the diagonal and $\varepsilon$ on the off-diagonal elements. 
If $K=1$ then $\mu=0.5$ and $\nu=5$.

We estimate the parameters using the variational EM for $K$ ranging from one to ten groups, and retain the solution maximizing the ICL as the optimal configuration overall.
As in the previous study, we compare the estimated maximum a posteriori partition with the true one using the ARI. 
The right panel of Figure \ref{fig:sim_1a_2a} shows the summaries for the ARI values in each of the five cases considered.
The index deviates from one as the number of groups increases, since the clustering and model choice tasks become more challenging.

In terms of model choice, table \ref{tab:sim_2a} reports the estimated and true values of $K$ through a confusion matrix.
\begin{table}[htb]
\centering
\begin{tabular}{c|cccccccccc}  
\toprule
&\multicolumn{10}{c}{$\hat{K}$} \\
\cmidrule(r){2-11}
& 1 & 2 & 3 & 4 & 5 & 6 & 7 & 8 & 9 & 10 \\
\midrule
$K = 1$ & {\color{red}1} & 0.00 & 0.00 & 0.00 & 0.00 & 0.00 & 0.00 & 0.00 & 0.00 & 0 \\
$K = 2$ & 0 & {\color{red}0.77} & 0.19 & 0.03 & 0.01 & 0.00 & 0.00 & 0.00 & 0.00 & 0 \\
$K = 3$ & 0 & 0.08 & {\color{red}0.62} & 0.18 & 0.06 & 0.04 & 0.02 & 0.00 & 0.00 & 0 \\
$K = 4$ & 0 & 0.02 & 0.25 & {\color{red}0.55} & 0.14 & 0.03 & 0.01 & 0.00 & 0.00 & 0 \\
$K = 5$ & 0 & 0.01 & 0.16 & 0.56 & {\color{red}0.15} & 0.08 & 0.02 & 0.01 & 0.01 & 0 \\
\bottomrule
\end{tabular}
\caption{\textbf{Simulation study 2}. Confusion matrix for the true value of $K$ versus its estimate $\hat{K}$. The values correspond to the proportion of generated datasets. The entries colored in red correspond to the proportion of datasets where the number of latent groups was correctly estimated.}
\label{tab:sim_2a}
\end{table}
The criterion performs very well when there are few or no groups, compared to the number of nodes. 
This is reasonable, since, if few clusters are present, more nodes will belong to the same group, thus highlighting the heterogeneity in the data.
By contrast, when the $100$ nodes are divided in $5$ different clusters, the algorithm struggles to recover the true partition exactly and tends towards underestimation of the actual number of groups. 
However, as shown in the right panel of Figure \ref{fig:sim_1a_2a}, the ARI scores are rather high in all datasets, signalling that the optimal clustering obtained is fundamentally similar to the data-generating one.

\subsection{Simulation study 3}
In the third simulation study we consider a more challenging situation where the time interval can be short.
This scenario poses a challenge since the available data that is used to estimate the model is rather limited.
Similarly to the previous simulation studies, $N=100$ networks are generated at random, using $K=3$ latent groups. In this study, the matrix $\bmu$ has values $\varepsilon = 0.5$ on the diagonal and $\theta = 5$ on the off-diagonal elements, and $\bnu$ has values $\theta$ on the diagonal and $\varepsilon$ on the off-diagonal elements. Therefore, for two nodes in the same group, the average interaction length is $5$, and the average non-interaction length is $0.2$. On the other hand, for nodes in different groups, the average interaction length is $0.5$, and the average non-interaction length is $5$. 

\begin{figure}[tb]
\centering
\includegraphics[width=0.49\textwidth]{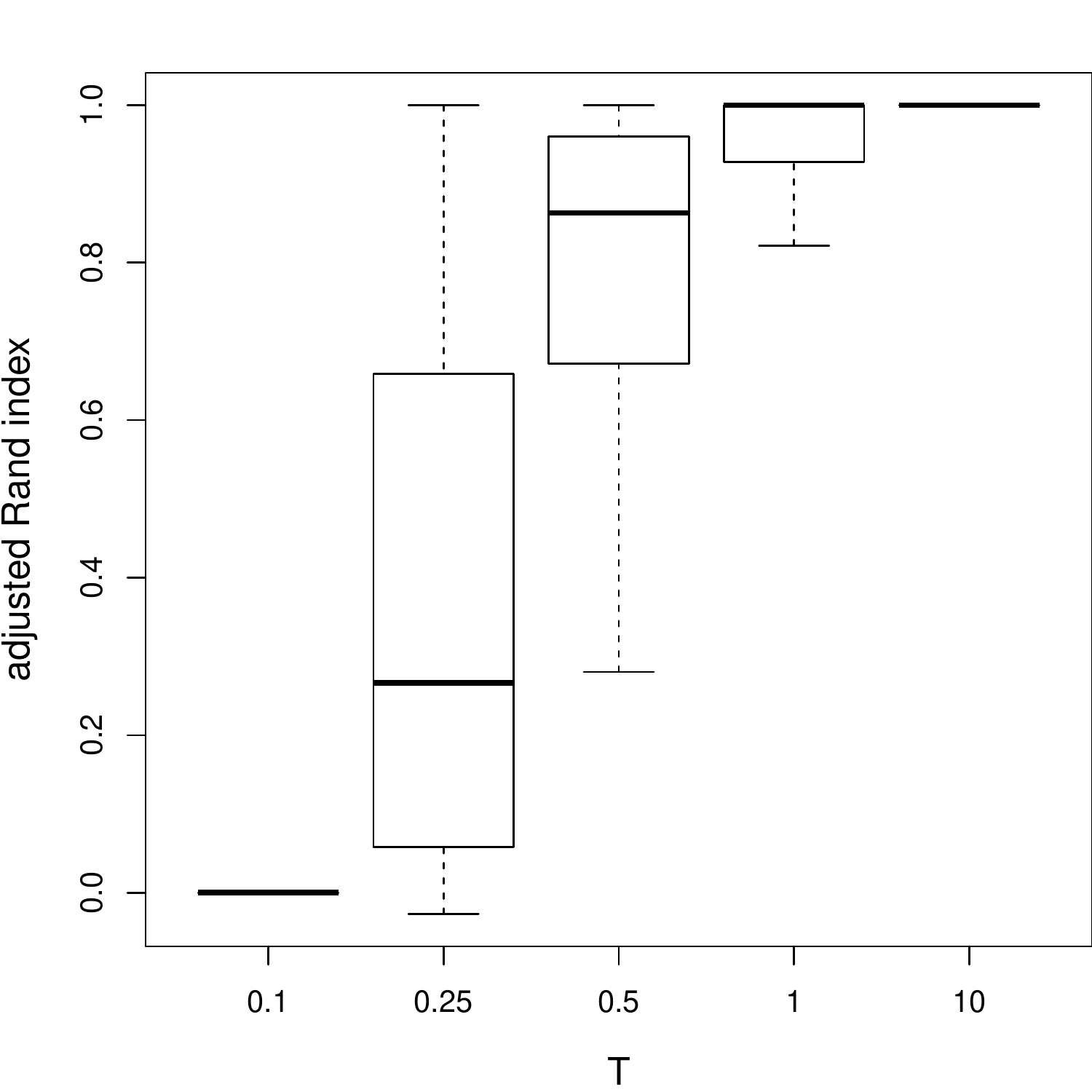}
\caption{\textbf{Simulation study 3}. Adjusted Rand index between the true partition and the estimated maximum a posteriori partition as a function of the interval length.}
 \label{fig:sim_3a}
\end{figure}

The time interval is denoted by $[0,T]$: we consider five different settings where $T$ varies in the set $\left\{0.1,0.25,0.5,1,10\right\}$, respectively.
Figure \ref{fig:sim_3a} illustrates the performance in each setting, measured by the ARI index, for the variational EM procedure. 
When the time interval is long enough, the algorithm achieves good results and recovers the latent structure without errors.
However, if $T$ is small, the available data is not sufficient to converge to a correct solution.
In particular, when $T=0.1$, most edges do not exhibit any change in time, meaning that the only interaction (or non-interaction) that may appear gets truncated on both sides.
The procedure manages to extract relevant information from the data as $T$ increases, especially for values larger than $1/\theta$, since, on average, edges will tend to exhibit at least one value change.

\section{High school students interaction data analysis}\label{sec:data}
In this section we show the proposed model in application to a dataset presented by \textcite{mastrandrea:2015}. The data concern face-to-face interactions among $327$ high school students in Marseilles, France, and were collected by means of wearable sensors over a period of $5$ days in December 2013. 
Students wore a sensor badge on their chest and the instrument recorded when they were facing each other with a time resolution of 20 seconds. 
Thus, any pair of students was considered interacting face-to-face when the sensors of the two were exchanging data packets at any given time during the $20$ seconds interval.

\begin{figure}[tb]
\centering
\includegraphics[scale=0.5]{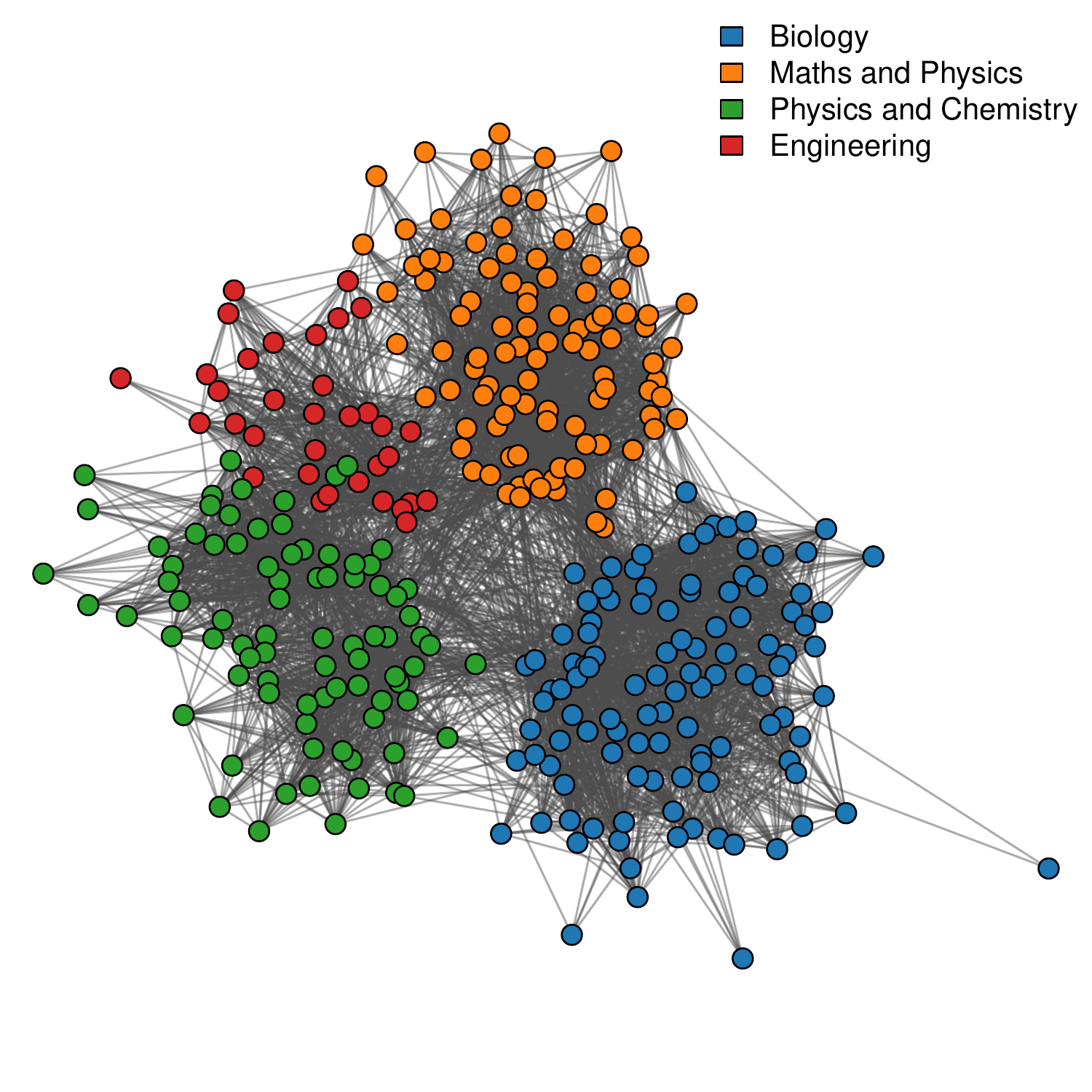}
\caption{\textbf{Data analysis}. Network of binary interactions among the high school students. Colors denote the different specializations.}
 \label{fig:net_progr}
\end{figure}

Additional information on the students is available from the same dataset. 
Students may have 4 different main specializations: biology (BIO), mathematics and physics (MP), physics and chemistry (PC), and engineering studies (PSI). Figure~\ref{fig:net_progr} shows a time-aggregated summary of the data through a binary interaction network between the students, i.e. an edge denotes that a pair of students had at least a single face-to-face contact during the $5$ days. 

Before using our proposed method, we need to transform the data to ensure that they take the form of a list of exponential lengths. Unfortunately, most data collected with wearable sensors are only available in a discrete representation, in fact, in our application the surroundings of an individual are scanned at $20$ seconds intervals.

We aim to transform the available data into a continuous collection of adjacency matrices $\bE$.
In order to do this, we focus our attention only on pairs of nodes that have a proximity contact more than $5$ times during the $5$ days.
For each of these pairs, we analyze the cadence of the contacts.
We start exploring their contact history from time $0$, and, whenever we encounter $5$ consecutive face-to-face contacts that take place in less than $5$ minutes, we start recording an interaction between the two nodes.
We continue browsing their history, and, as soon as $5$ consecutive face-to-face contacts extend to more than $5$ minutes, we interrupt the interaction between the nodes.
In this way, for every pair of nodes, we construct interactions of variable length, which appear in those time sections where face-to-face contacts between the two students were more frequent.
In fact, one feature of this construction is that the start time and end time of interactions coincide with face-to-face contact times.
We note that this transformation shifts the interactions in time, but this has no effect on our analysis, since we only focus on the length of the interactions, and not necessarily on their positioning.

Our goal now is to compare the clustering structure informed by the constructed interaction lengths to the four different specializations that are observed.
Hence we estimate the proposed SBM by running the algorithm with $K$ fixed to 4 in advance. 

\begin{figure}[ht]
\centering
\includegraphics[scale=0.5]{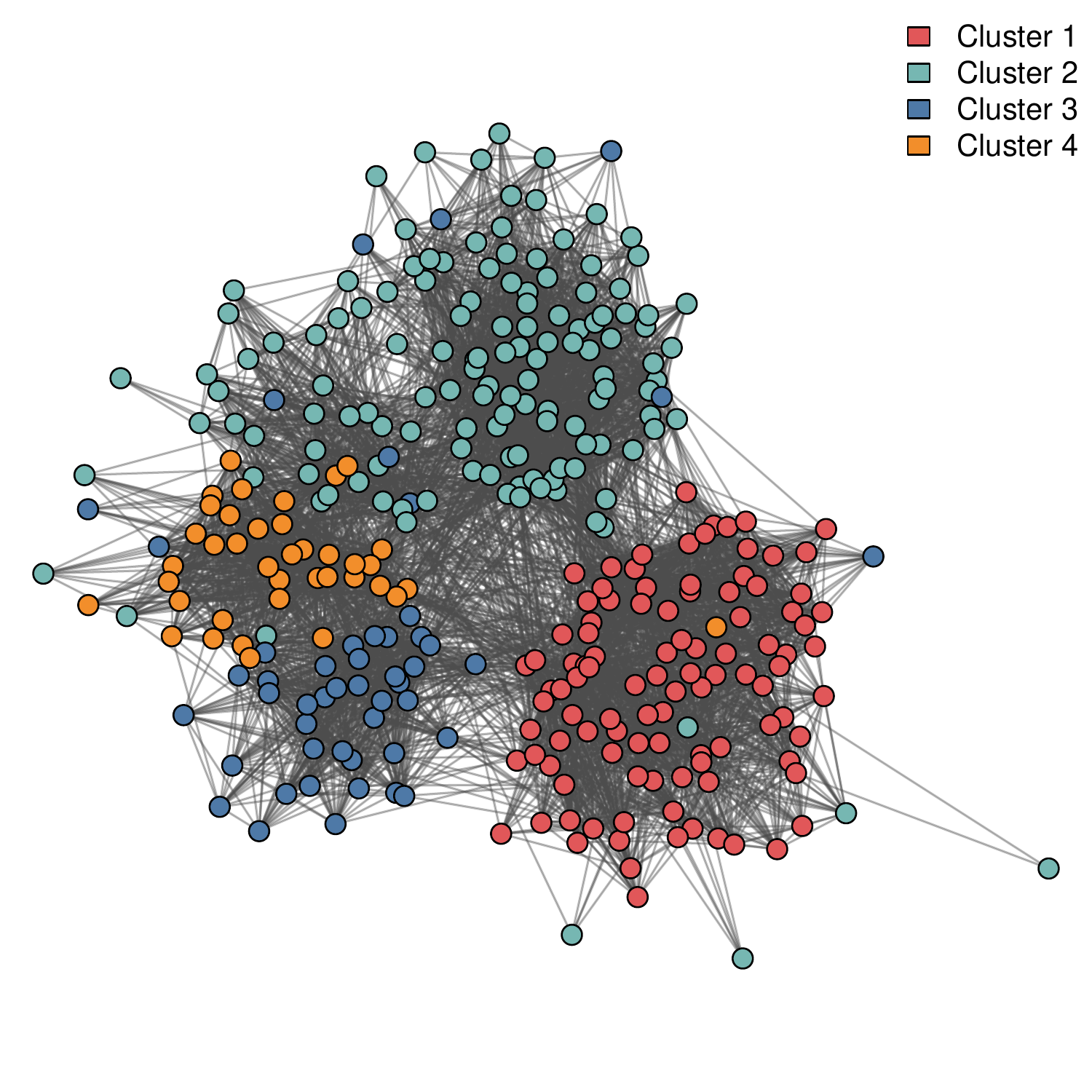}\qquad
\includegraphics[scale=0.5]{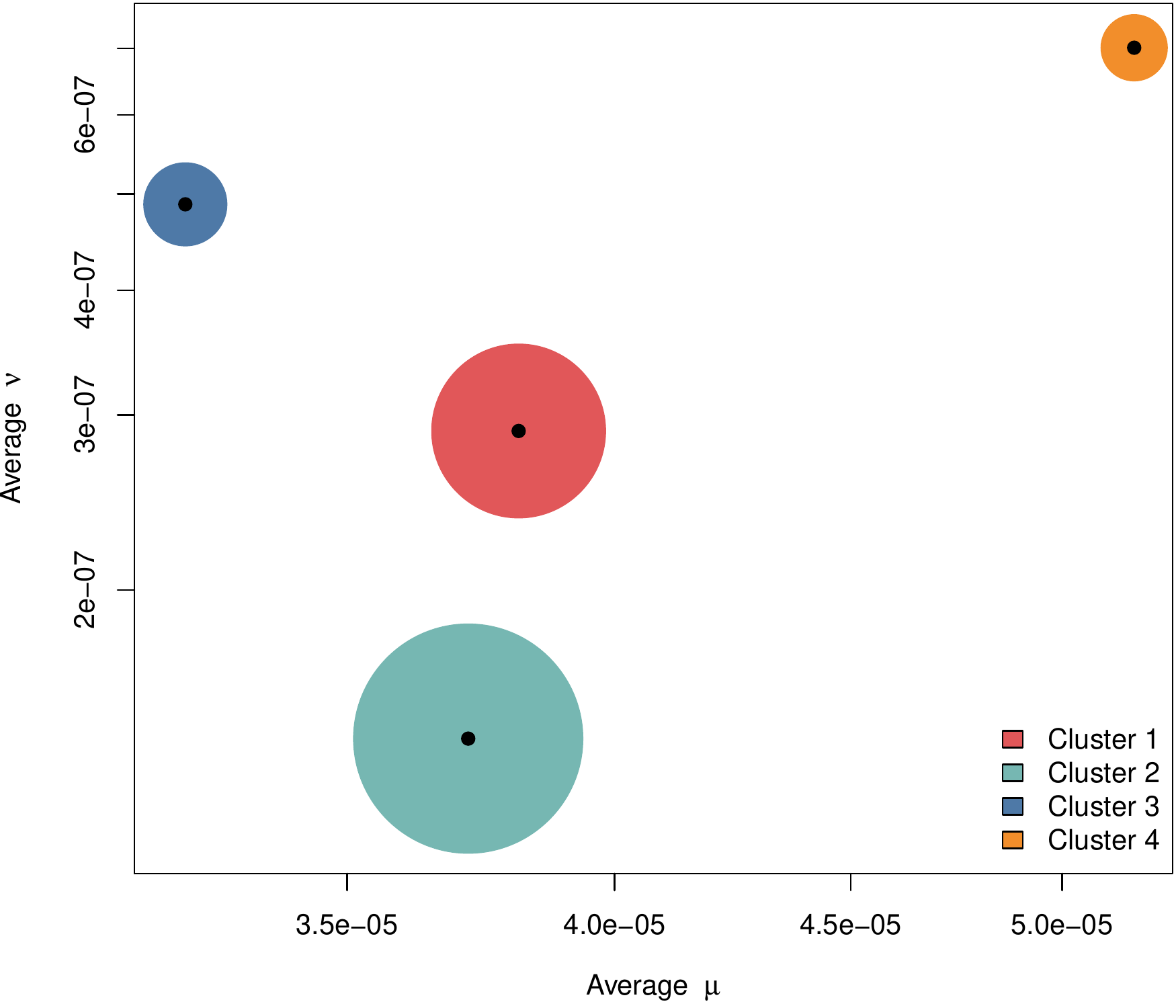}
\caption{\textbf{Data analysis}. Left: Network of binary interactions among the high school students; nodes are colored according to the inferred clustering. Right: Average values of the estimated interaction length parameters $\bmu$ versus the average values of the estimated non-interaction length parameters $\bnu$ (logarithmic scale); size of the circle is proportional to the size of the corresponding cluster.}
 \label{fig:net_clust}
\end{figure}

The results are presented in the confusion matrix in Table~\ref{tab:high_school} and in Figure~\ref{fig:net_clust}. 
\begin{table}[bt]
\centering
\begin{tabular}{llrrrr}
\toprule
&  & \multicolumn{4}{c}{\em Cluster}\\
& & 1 & 2 & 3 & 4 \\ 
\midrule
\multirow{4}{*}{\em Specialization} & BIO & 103 &   5 &   1 &   1 \\ 
& MP &    &  96 &   4 &    \\ 
& PC &    &   4 &  41 &  38 \\ 
& PSI &    &  31 &   3 &    \\ 
\bottomrule
\end{tabular}
\caption{\textbf{Data analysis}. Confusion matrix among the classification of students according their major specialization and the estimated classification.}
\label{tab:high_school}
\end{table}
The ARI between the estimated partition and the different specialization classes is equal to 0.66, denoting good correspondence. From the table, Cluster 1 only corresponds to students in biology, while Cluster 2 is mainly a mixture of students taking the specialization in mathematics and physics and the specialization in engineering. The class corresponding to students in physics and chemistry is split among Cluster 3 and 4. Interestingly, these students are separated into two blocks, characterized by very dissimilar average interaction lengths. The right panel of  Figure~\ref{fig:net_clust} reports the average values of the estimated interaction length parameters $\bmu$ versus the average values of the estimated non-interaction length parameters $\bnu$ (logarithmic scale). Cluster 4 has a smaller average time of interaction compared to the others and particularly to Cluster 3, which is the one with the longest average interaction time. Moreover, clusters are distinguished also in terms of average non-interaction time, with units in Cluster 1 being those not having face-to-face contacts for longer periods on average.  

We note that, in figures \ref{fig:net_progr} and \ref{fig:net_clust}, the positions of the nodes are deduced from their binary interactions. This means that the lengths of interactions are not used in any way, hence, these graphical representations may not necessarily exhibit the true underlying topology of the observed network.
For example, differently from the given clustering configuration, our method allocates several nodes that are positioned in the outskirts of the network (Figure \ref{fig:net_clust}, left panel) into cluster $2$. 
From a static SBM point of view, the behavior of these nodes is not reasonable, since it does not align with the exhibited block and community structure.
In other words, in the static SBM context, these nodes add a lot of heterogeneity to cluster $2$.
On the other hand, the goal of our method is to cluster together nodes that have similar interaction lengths.
This means that, while nodes in cluster $2$ have a heterogeneous behavior when choosing their neighbors, they also tend to create connections of similar lengths.
Based on the right panel of Figure \ref{fig:net_clust}, we can state that the nodes in cluster $2$ create both long interactions and long non-interactions.


\section{Conclusions}\label{sec:conclusions}
The model introduced in this paper provides a new approach to analyze networks evolving over time. 
The main advantage of the proposed model is its fully continuous specification, which allows more flexibility and, possibly, a better fit to the observed data.
Thanks to the SBM structure, the estimation of the model parameters can be performed efficiently with an adapted EM algorithm.
We have shown this estimation method to be effective in our simulation studies, where the procedure successfully recovered the latent structure in a variety of scenarios.
We have also proposed an adaptation of the integrated completed likelihood criterion, showing that it generally allows one to discover the number of latent clusters.

Since most observed networks evolve in a continuous fashion over time, the observed initial and final interaction lengths are generally truncated, due to the arbitrary data collection method that is used.
Our approach includes this property directly in the modelling, hence leading to a formulation that is theoretically appropriate also in cases where the time interval is short.

This methodology can be used for both directed and undirected interactions, and it may be extended to a bipartite network context. 
The model may also be extended to allow the parameters $\bmu$ and $\bnu$ to change over time.
This type of idea has been recently explored for other types of blockmodels, see \textcite{matias2018semiparametric} and \textcite{corneli2017multiple}.
Our same model may also be considered in a conjugate Bayesian setting, since conjugate priors may be specified for all of the model parameters.

One limitation of the approach proposed ensues from the multimodality of the objective function.
While the variational objective is generally smoother than the actual marginal likelihood of the model, the algorithm adopted remains a heuristic one, in that there are no guarantees that the optimal solution found corresponds to the global maximum.
This fact is a known hindrance in SBMs, and particularly in dynamic SBMs.
The initialization method described in this paper may help prevent the local optimum issue, but it does not solve this problem. 

Future work may build upon our proposal both in terms of modelling and on the computational aspects, by defining new continuous dynamic blockmodel frameworks, and by considering more effective estimation procedures, respectively.

\section*{Software}
The \texttt{R} package \texttt{expSBM} accompanies this paper and it provides an implementation of the variational algorithm described, for both directed and undirected networks. 
Parts of the code have been written in \texttt{C++} to reduce the overall computing time.
The package is publicly available from \texttt{CRAN} \parencite{rcoreteam}.


\printbibliography

\newpage

\footnotesize
\appendix
\section{Appendix}
\footnotesize

\subsection{Proof of Proposition \ref{prop:elbo_1}}\label{app:elbo_1}
The evidence lower bound is defined as follows:
\begin{equation}
    \begin{split}
        \mathcal{F} 
        &= \mathbb{E}_q\left[ \log p\left( \bE, \textbf{Z} \middle \vert \bmu, \bnu,  \blambda \right) \right] + Ent(q) 
        = \mathbb{E}_q\left[ \ell_{\bE}\left( \bmu, \bnu, \textbf{Z} \right)  \right] + \mathbb{E}_q\left[ \log p\left( \textbf{Z} \middle \vert \blambda \right) \right] - \mathbb{E}_q\left[ \log q\left( \textbf{Z} \middle\vert \btau \right) \right] \\
    \end{split}
\end{equation}
We study the terms on the right hand side separately.
\begin{equation}
    \begin{split}
        \mathbb{E}_q\left[ \ell_{\bE}\left( \bmu, \bnu, \textbf{Z} \right)  \right]
        &= \mathbb{E}_q\left[\sum_{g=1}^K \sum_{h=1}^K \left\{L_{\mu_{gh}} \log \left( \mu_{gh} \right) + L_{\nu_{gh}} \log \left( \nu_{gh} \right) - \mu_{gh}\eta_{gh} - \nu_{gh}\zeta_{gh}\right\} \right] \\
        &= \sum_{g=1}^K \sum_{h=1}^K \left\{\mathbb{E}_q\left[L_{\mu_{gh}}\right] \log \left( \mu_{gh} \right) 
        + \mathbb{E}_q\left[L_{\nu_{gh}}\right] \log \left( \nu_{gh} \right) 
        - \mu_{gh}\mathbb{E}_q\left[\eta_{gh}\right] 
        - \nu_{gh}\mathbb{E}_q\left[\zeta_{gh} \right] \right\} \\
        &= \sum_{g=1}^K \sum_{h=1}^K \left\{ \left(\sum_{i \neq j} \tau_{ig} \tau_{jh} \mathcal{A}_{ij}^{(+)}\right) \log \left( \mu_{gh} \right) 
        + \left(\sum_{i \neq j} \tau_{ig} \tau_{jh} \mathcal{A}_{ij}^{(-)}\right) \log \left( \nu_{gh} \right) \right. \\
        &\hspace{1cm}- \left.\left(\sum_{i \neq j} \tau_{ig} \tau_{jh} \mathcal{X}_{ij}^{(+)}\right) \mu_{gh} 
        - \left(\sum_{i \neq j} \tau_{ig} \tau_{jh} \mathcal{X}_{ij}^{(-)}\right) \nu_{gh} \right\} \\
        &= \sum_{g=1}^K \sum_{h=1}^K \left\{\bar{L}_{\mu_{gh}} \log \left( \mu_{gh} \right) + \bar{L}_{\nu_{gh}} \log \left( \nu_{gh} \right) - \mu_{gh}\bar{\eta}_{gh} - \nu_{gh}\bar{\zeta}_{gh}\right\}
    \end{split}
\end{equation}
\begin{equation}
    \begin{split}
        \mathbb{E}_q\left[ \log p\left( \textbf{Z} \middle \vert \blambda \right) \right]
        &= \mathbb{E}_q\left[ \sum_{k=1}^K\sum_{i=1}^N Z_{ik}\log \lambda_k\right] 
        = \sum_{k=1}^K\sum_{i=1}^N\mathbb{E}_q\left[  Z_{ik}\right]\log \lambda_k
        = \sum_{k=1}^K\sum_{i=1}^N\tau_{ik}\log \lambda_k
    \end{split}
\end{equation}
\begin{equation}
    \begin{split}
        \mathbb{E}_q\left[ \log q\left( \textbf{Z} \middle \vert \btau \right) \right]
        &= \mathbb{E}_q\left[ \sum_{k=1}^K\sum_{i=1}^N Z_{ik}\log \tau_{ik}\right] 
        = \sum_{k=1}^K\sum_{i=1}^N\tau_{ik}\log \tau_{ik}
    \end{split}
\end{equation}
The three parts combined give \eqref{prop:elbo_1_eq_1}.

\subsection{Proof of Proposition \ref{prop:elbo_tau}}\label{app:elbo_tau}
The evidence lower bound can be rewritten as follows:
\begin{equation}
    \begin{split}
        \mathcal{F}
        &= \sum_{g=1}^K \sum_{h=1}^K \left\{ 
        \left( \sum_{i \neq j} \tau_{ig}\tau_{jh} \mathcal{A}_{ij}^{(+)}\right) \log \mu_{gh}
        + \left( \sum_{i \neq j} \tau_{ig}\tau_{jh} \mathcal{A}_{ij}^{(-)}\right) \log \nu_{gh} \right. \\
        &\hspace{1cm} \left. - \left( \sum_{i \neq j} \tau_{ig}\tau_{jh} \mathcal{X}_{ij}^{(+)}\right) \mu_{gh}
        - \left( \sum_{i \neq j} \tau_{ig}\tau_{jh} \mathcal{X}_{ij}^{(-)}\right) \nu_{gh}
        \right\}
        + \sum_{i=1}^N \sum_{k=1}^K \tau_{ik} \log \lambda_k + \sum_{i=1}^N \sum_{k=1}^K \tau_{ik} \log \tau_{ik} \\
        &= \sum_{g=1}^K \sum_{h=1}^K \sum_{i \neq j} \tau_{ig}\tau_{jh} \omega_{ijgh}
        + \sum_{i=1}^N \sum_{k=1}^K \tau_{ik} \log \lambda_k + \sum_{i=1}^N \sum_{k=1}^K \tau_{ik} \log \tau_{ik} \\
    \end{split}
\end{equation}
Now consider the following Lagrangian:
\begin{equation}
    \mathcal{H} = \mathcal{F} + \sum_{i=1}^N \xi_{i} \left( \sum_{k=1}^K \tau_{ik}-1 \right)
\end{equation}
with multipliers $\xi_1,\dots,\xi_N$.
The derivative is equal to the following:
\begin{equation}
    \frac{\partial \mathcal{H}}{\partial \tau_{\ell k}}
    = \sum_{j=1}^N \sum_{h=1}^K \tau_{jh} \omega_{\ell jkh}
    + \sum_{i=1}^N \sum_{g=1}^K \tau_{ig} \omega_{i\ell gk}
    + \log \lambda_k - \log \tau_{\ell k} - 1 + \xi_{\ell}
\end{equation}
with root:
\begin{equation}\label{prop:elbo_tau_eq_3}
    \tau_{\ell k} = \exp\left\{ \sum_{j=1}^N \sum_{h=1}^K \tau_{jh} \omega_{\ell jkh}
    + \sum_{i=1}^N \sum_{g=1}^K \tau_{ig} \omega_{i\ell gk}
    + \log \lambda_k - 1 + \xi_{\ell}\right\}
\end{equation}
Regarding the constraints:
\begin{equation}
    1 = \sum_{k=1}^K \tau_{\ell k} = \exp\left\{\xi_{\ell}\right\} \sum_{k=1}^K \exp\left\{ \sum_{j=1}^N \sum_{h=1}^K \tau_{jh} \omega_{\ell jkh}
    + \sum_{i=1}^N \sum_{g=1}^K \tau_{ig} \omega_{i\ell gk}
    + \log \lambda_k - 1\right\}
\end{equation}
This yields the following:
\begin{equation}
    \xi_{\ell} = \log \sum_{k=1}^K \exp\left\{ \sum_{j=1}^N \sum_{h=1}^K \tau_{jh} \omega_{\ell jkh}
    + \sum_{i=1}^N \sum_{g=1}^K \tau_{ig} \omega_{i\ell gk}
    + \log \lambda_k - 1\right\}
\end{equation}
This critical point is a maximum.
Using this result in \eqref{prop:elbo_tau_eq_3} finishes the proof.

\subsection{Proof of Proposition \ref{prop:elbo_lambda}}\label{app:elbo_lambda}
Consider the following Lagrangian:
\begin{equation}
    \mathcal{H} = \sum_{g=1}^K \left(\sum_{i=1}^N \tau_{ig}\right) \log \lambda_g + \xi \left( \sum_{g=1}^K \lambda_g -1\right)
\end{equation}
and its derivative:
\begin{equation}
    \frac{\partial \mathcal{H}}{\partial \lambda_k}
    = \frac{\sum_{i=1}^N \tau_{ik}}{\lambda_k} + \xi
\end{equation}
This gives the root $\lambda_k = - \sum_{i=1}^N \tau_{ik} / \xi$ and in turn:
\begin{equation}
    \xi = -\sum_{i=1}^N \sum_{g=1}^K \tau_{ig} = -N
\end{equation}
which leads to the result of the proposition.
This critical point is a maximum.

\subsection{Proof of Proposition \ref{prop:elbo_munu}}\label{app:elbo_munu}
From \eqref{prop:elbo_1_eq_1}:
\begin{equation}
    \frac{\partial \mathcal{F}}{\partial \mu_{gh}}
    = \frac{ \bar{L}_{\mu_{gh}} }{ \mu_{gh} } - \bar{\eta}_{gh}
\end{equation}
has root $\mu_{gh} =  \frac{ \bar{L}_{\mu_{gh}} }{ \bar{\eta}_{gh} }$ which corresponds to a maximum. 
The formula for $\nu_{gh}$ is obtained analogously.

\end{document}